\documentclass{article}
\AtBeginDocument{%
  \providecommand\BibTeX{{%
    \normalfont B\kern-0.5em{\scshape i\kern-0.25em b}\kern-0.8em\TeX}}}

\usepackage{enumitem}
\setlist[itemize]{listparindent=\parindent,align=parleft,leftmargin=0pt}
\usepackage{graphicx}    
\usepackage{amsmath}     
\usepackage{amsfonts}    
\usepackage{amssymb}     
\usepackage{hyperref}    
\usepackage{booktabs} 
\usepackage{natbib}
\usepackage{url,hyperref,lineno,microtype,subcaption}
\usepackage[onehalfspacing]{setspace}

\begin{document}

\title{Designing Interactions with Autonomous Physical Systems}

\maketitle
\begin{abstract}
  In this position paper, we present a collection of four different prototyping approaches which we have developed and applied to prototype and evaluate interfaces for and interactions around autonomous physical systems. Further, we provide a classification of our approaches aiming to support other researchers and designers in choosing appropriate prototyping platforms and representations.
\end{abstract}
\section{Introduction}
Ongoing advances in computing, sensing and network technologies lead to increasing deployment of autonomous physical systems in real-world urban contexts \cite{While2020}. 
However, a key to the successful uptake of these technologies is their public acceptance and trustworthiness \cite{Nordhoff2018}, which is also closely linked to usability and user experience (UX) \cite{Frison2019}. Therefore, the field of human-computer interaction (HCI) and more specifically research in Explainable Artificial Intelligence (XAI) can play an important role in the development of human-machine interfaces (HMI) that make the decisions made by autonomous physical systems transparent, interpretable and explainable \cite{Anjomshoae2019}. This has been partially investigated, for example, in the context of autonomous vehicles (AV), using external HMIs to communicate the system's awareness and intents to surrounding pedestrians \cite{Dey2020taming}. In addition, HCI offers methods and tools for prototyping and evaluating such interfaces following a human-centred approach \cite{Tomitsch2021}. This is in particular important as the viewpoint and expertise of a variety of stakeholders is required. However, testing is cumbersome and challenging due to the complexity of the design context and using actual prototypes for evaluations with users may put participants at risk. Therefore, researchers have to fall back on simulation platforms and low-risk prototyping approaches.

In this paper, we present a series of prototyping approaches which we have used for designing and evaluating interfaces for \textit{and} interactions around autonomous physical systems in urban environments (hereafter referred to as urban robots). We then introduce a preliminary classification of our approaches, aiming to support other HCI-researchers and practitioners to choose appropriate prototyping platforms and representations depending on the specific questions that they are seeking to investigate, the design process stage, and feasibility. The paper concludes with a research agenda on under-explored topics in explainable HMIs for autonomous physical systems, and how prototyping approaches need to be further adapted for their systematic investigation.



\begin{figure*}
\begin{center}
  \includegraphics[width=0.80\textwidth]{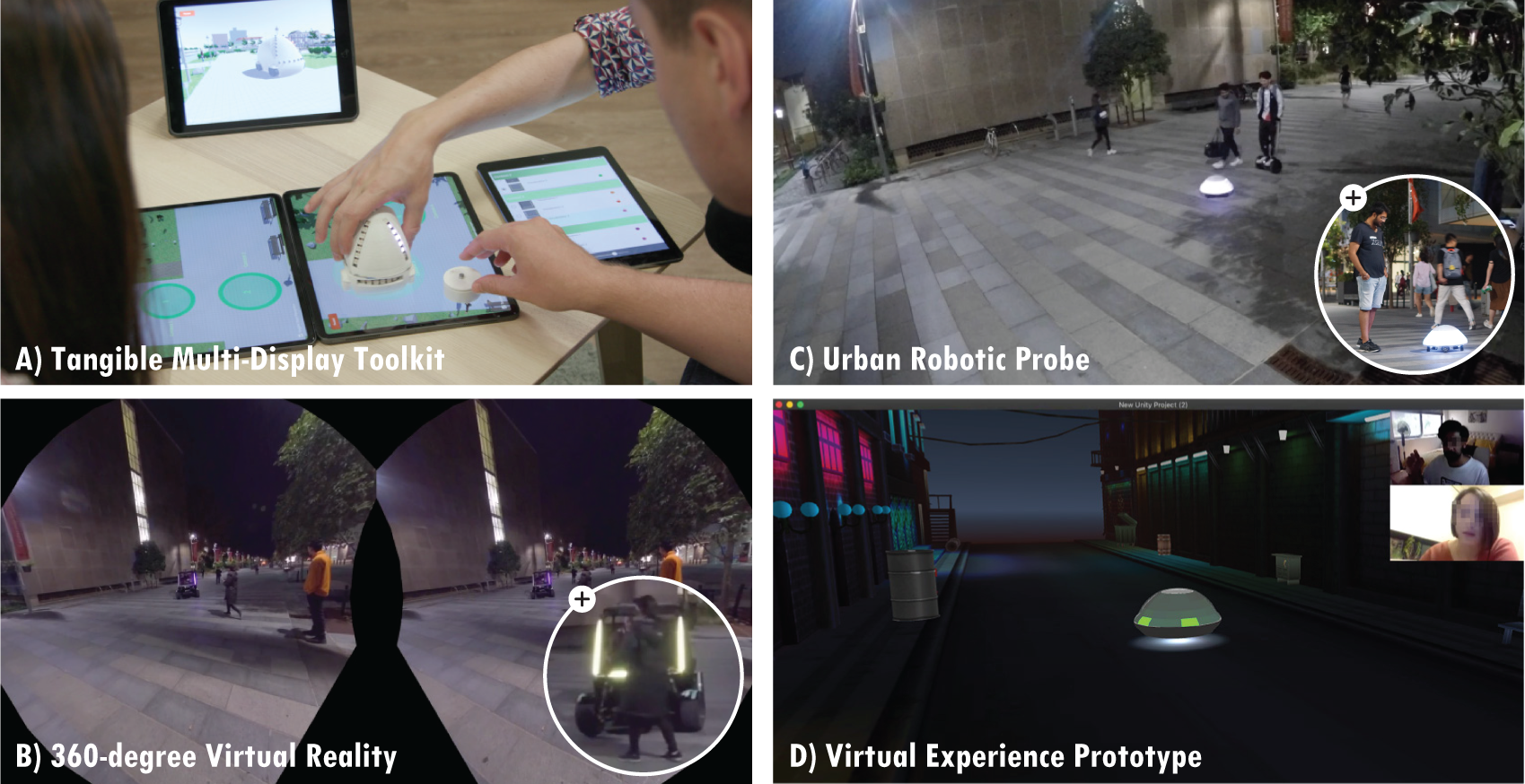}
  \caption{Four prototyping approaches for designing urban robotic interfaces.}
  \label{fig:teaser}
  \end{center}
\end{figure*}

\section{Prototyping Approaches}
Prototyping is an integral part of a human-centred design process \cite{Buxton2007}. Usually, designers start with low-fidelity approaches (e.g. paper prototypes), and later in the design process move on to prototyping at higher detail levels (e.g. mock-ups). 
Often the notion of fidelity is closely linked to the matter of cost, however researchers also emphasised the importance of a more distinct characterisation of prototypes and how it is linked to different qualities of interest \cite{Lim2008}. In the case of HMIs for urban robots, prototyping becomes even more challenging (compared for example to prototyping a website or mobile application) because: interface behaviours are closely linked to the situational context; in urban environments social interactions might be at play, and various user roles need to be considered simultaneously (i.e. people interacting in close proxemics with the system, casual bystanders \cite{Boll2019}).
In the following, we present four prototyping approaches (see Figure \ref{fig:teaser}, A-D) which we have applied in two large-scale research projects on urban robots. While these urban robots differ in scale, application type, interface qualities, and the specific research questions we aimed to address, they have in common an HMI that enables them to communicate with surrounding people. 

\begin{itemize}[label={}]
\item \textbf{A) Tangible Multi-Display Toolkit:} We developed a toolkit approach for AV-pedestrian interface design explorations that enables multiple viewing angles and perspectives to be captured simultaneously (e.g. top-view, first-person pedestrian view) through computer-generated imagery orchestrated across multiple displays \cite{Hoggenmueller2020ozchi}. Users are able to directly interact with the simulated environment through tangibles, which physically simulate the interface’s behaviour (e.g. through an integrated LED display). We used the small-scale scenario prototype in collaborative sessions with experts to collect early feedback on the design of an AV-pedestrian interface for a driver-less electric pod.
\item \textbf{B) 360-degree Virtual Reality:} While the majority of studies on AV-pedestrian interfaces implement computer-generated VR, 360-degree VR can simulate the experience of interacting with real-world prototypes at high visual realism. After collecting initial feedback through approach A, we developed a low-resolution lighting display attached to an actual pod. We used a 360-degree camera to video record different scenarios in a shared space environment (e.g. AV picking up another pedestrian). After deploying the videos into the game development platform Unity\footnote{\url{https://unity.com/}, last accessed: April 2021}, we evaluated the HMI with users in immersive VR, thereby measuring trust and UX \cite{Hoggenmueller2020chi}.
\item \textbf{C) Urban Robotic Probe:} We applied the concept of urban probes - light-weight prototypes to collect feedback about the use of urban technologies in real-world settings \cite{Paulos2005} - to the context of urban robotic experimentation. We designed and developed a speculative urban robot application using a self-moving, autonomously powered platform \cite{Hoggenmueller2021chi}. We deployed the prototype in a series of in-the-wild studies: for example, to evaluate affective expressions through movement and light-patterns with the overall aim to increase the social acceptance of urban robots \cite{Hoggenmueller2020perdis}.
\item \textbf{D) Virtual Experience Prototypes:} Building on the concept of experience prototyping \cite{Buchenau2000}, we investigated how to gather qualitative feedback in an online evaluation context through Virtual Experience Prototypes (VEPs) \cite{Almo2020ozchi}. For our study, we created a non-immersive VEP consisting of a 3D-model of the existing physical prototype and a virtual urban environment similar to the initial deployment location in C. We used the VEP to inform the design of robotic expressions (i.e. through light and sound) in virtually conducted expert workshops. We collected qualitative data through the think-aloud protocol while participants interacted with the VEP, followed by subsequent in-depth interviews.
\end{itemize}

\begin{figure*}[b]
\begin{center}
  \includegraphics[width=\textwidth]{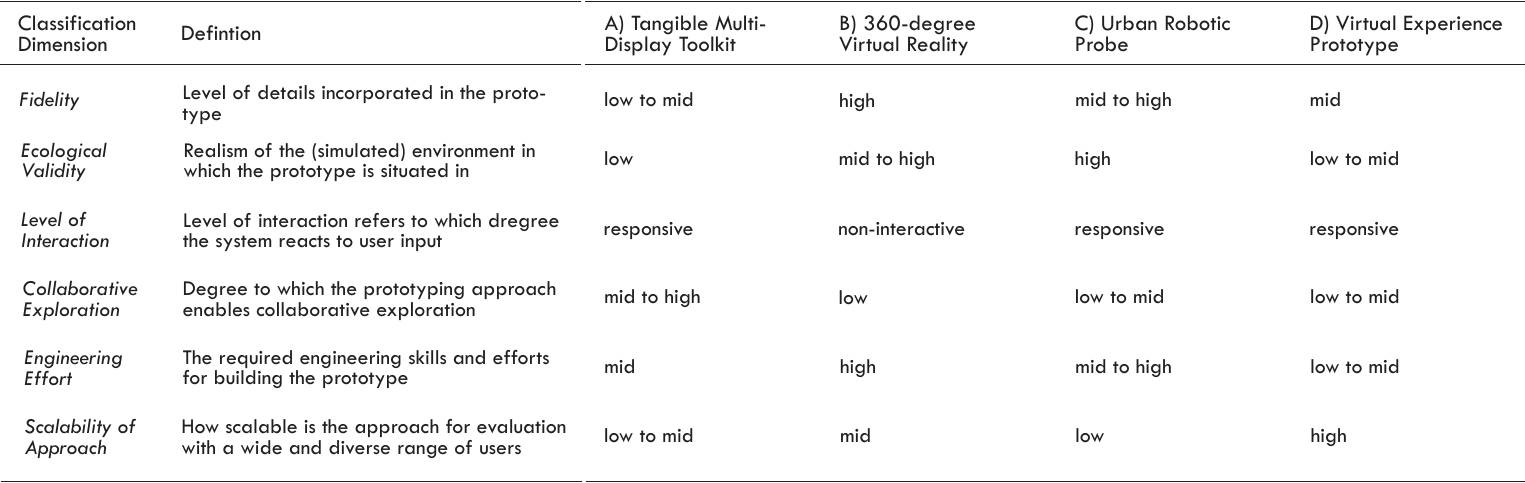}
  \caption{Classification of the prototyping approaches and definition of the classification dimensions.}
  \label{fig:classification}
  \end{center}
\end{figure*}

\section{Classification}

\noindent In Table \ref{fig:classification}, we introduce a preliminary classification of our prototyping approaches, thereby drawing on a classification system inspired by \cite{wiethoff2012} and \cite{Lim2008}.

\begin{itemize}[label={}]
\item \textit{Fidelity:} The notion of fidelity (probably the most common for the characterisation of prototypes) refers to the level of details incorporated in the robotic artefacts. Approach A and D are situated in the low to mid spectrum as they incorporate some level of details (e.g. dynamic interface content), which is not manifested in paper prototypes. We rank D slightly higher as also the robot's movement can be simulated. B and C are situated on the higher end of the fidelity spectrum: here, we built on two fully-functional robotic platforms, with the prototype in B even moving fully autonomously based on sensor input.

\item \textit{Ecological Validity:} For context-based interfaces, the realism of the (simulated) environment is important, which we incorporate in the notion of ecological validity. While our robotic probe in C is lower in fidelity compared to the pod in B, the representation through VR (B) can not convey all experiential qualities of a real-world deployment (C). Nevertheless, we consider both approaches as valuable when it comes to a more holistic evaluation of urban robots, thereby also considering spatial and contextual aspects.

\item \textit{Level of Interaction:} With the notion of interactivity, we refer to three categories (non-interactive, responsive, interactive) to classify to which degree the prototypical HMI reacts to user input. Three of our approaches (A, C, D) can simulate responsive interface behaviour through various user input. For example, in C we tested ultrasonic sensors to adapt the robot's light patterns depending on the user's proximity. For D, responsive behaviour was simulated through the user moving in the environment using keyboard buttons. In this regard, a disadvantage of using 360-degree Virtual Reality (B) becomes apparent where responsive and interactive behaviour can only be demonstrated through the recording of staged interactions, but not actively experienced by the VR participant itself.

\item \textit{Collaborative Exploration:} Designing urban robots for real-world contexts requires the involvement of a wide range of stakeholders, including engineers, interaction designers, urban planners and citizens. Therefore, providing prototyping platforms to foster collaborative explorations is necessary. While A ranks low in fidelity and ecological validity, the approach is in particular well suited for collaborative explorations. We were able to use the toolkit to explore interface concepts in workshop sessions, thereby also incorporating existing brainstorming materials (e.g. post-its). 

\item \textit{Engineering Effort:} This dimension refers to the burdens to develop the prototype in terms of time, cost and skills required. For our approaches, the engineering effort mostly correlates with the level of fidelity. An exception here is approach D, where mid-fidelity levels were reached through relatively low efforts by using the game development platform Unity, which comes with a range of tutorials and an active online community. 

\item \textit{Scalability of Approach:} Here, we refer to the scalability of the approach for evaluating the prototype with a large and diverse range of users. In this regard, D exceeds all other approaches as evaluations are possible without requiring specialised hardware while reaching out to a large pool of users.
\end{itemize}

The various dimensions introduced in this classification play a different role depending on the design process stage and the specific research questions under investigation. For example in an early stage, it is important that the approach enables to quickly generate and evaluate a wide range of interface concepts, thereby seeking to reach out to a wide range of stakeholders bringing in various expertise. Therefore, fidelity and ecological validity, we argue, are less important compared to enabling collaborative explorations and a low engineering effort. On the other hand, when it comes to generate measurable outcomes (e.g. how does the HMI supports safety and trust), it is more important that user's perception and behaviour are comparable to a real-world situation \cite{FELDSTEIN2020105356}. In this regard, we also found in a comparison study of 360-degree VR (B) and computer-generated VR simulations that differences in spatial awareness and perceived realism influence on which experiential and perceptual aspects user focus in the evaluation \cite{Hoggenmueller2021chi}. We found that when evaluating trust and UX, aspects beyond the AV and the eHMI itself are assessed, for example overall trust towards the situation (e.g. surrounding environment and other people). Here, the prototype representation in 360-degree VR was considered as more ``trustworthy'' due to the higher realism and more natural depictions of people interacting with the AV. When it comes to the evaluation in real-world urban contexts, we think that low-risk applications such as the urban robot in C could further serve as a proxy to test urban interactions between passers-by and autonomous physical systems in a quick and safe manner. Here, future research would need to investigate to which degree the insights generated through light-weight urban robotic probes are also applicable to high-risk applications, such as pods.

\section{Workshop Contribution and Future Work}
By presenting various prototyping approaches and a preliminary classification of those approaches, we hope to contribute to the larger research context of explainable and trustworthy autonomous physical systems through supporting other researchers in choosing appropriate prototyping platforms and representations. We argue that for the successful deployment of autonomous physical systems, it is key to prototype and test those systems at various stages of the design process following a human-centred approach \cite{Tomitsch2021}. 
Building on previous research in HCI that emphasised the importance of understanding the fundamental characteristics of prototypes (e.g. paper prototypes) in the context of interactive products \cite{Lim2008}, we consider our investigation as a first step to understand the characteristics of emerging prototyping platforms (e.g. VR) which are increasingly being used to prototype interfaces for autonomous physical systems.

Previous research has investigated XAI mostly in data-driven contexts and applications \cite{Anjomshoae2019}. We agree with the workshop organisers that it is timely to investigate the aspects of interpretability and explainability also in the context of autonomous physical systems, given their increasing deployment in real-world urban contexts \cite{McGuirk2020}. In the case of AVs, research studies and commercial applications address this challenge for in-car passengers through providing a visualisation of what the system is able to see based on its sensors, and give them the ability to intervene if necessary \cite{Kauer2010}. It is, however, much more challenging to address this for people in the surrounding environment, who may not even be aware that they take part in this interaction \cite{Boll2019}. Recent work within HCI, including our own work presented in this paper, has studied this challenge for AVs operating on roads, using HMIs to communicate the system's awareness and intent. While these HMI concepts are designed to communicate an AV’s internal state, it remains difficult for people around the vehicle to understand how a particular decision is made. In our future research, we therefore aim to develop HMI concepts that communicate to surrounding pedestrians the decision-making process itself. In this regard, we think that also the aspect of interactivity is mostly untapped as current HMIs are designed to broadcast information and do not offer people means to directly influence their behaviour. With regard to our here presented prototyping approaches, this also indicates that further development is necessary to support prototyping and evaluating multi-user interfaces to enable interactions between pedestrians and autonomous physical systems.

\bibliographystyle{ACM-Reference-Format}
\bibliography{sample-base}

\end{document}